\documentclass[aps,prl,twocolumn]{revtex4}

\usepackage{graphicx}
\usepackage{amsmath}
\usepackage{bm}
\usepackage{bbold}
\usepackage[dvips]{color}

\newcommand{\be}{\begin{equation}}
\newcommand{\ee}{\end{equation}}
\newcommand{\bwt}{\begin{widetext}}
\newcommand{\ewt}{\end{widetext}}
\newcommand{\bea}{\begin{eqnarray}}
\newcommand{\eea}{\end{eqnarray}}

\begin{document}
\title{Maintaining Quantum Coherence in the Presence of Noise by
  State Monitoring}

\author{T.~Konrad,$^{1,2}$\protect\footnote{Electronic address:
\textit{konradt@ukzn.ac.za}}, H.~Uys$^{3}$\protect\footnote{Electronic address:
\textit{huys@csir.co.za}} }
\affiliation{\mbox{$^1$ Center for Quantum Technologies, School of
    Physics,  University of KwaZulu, Natal, Durban, South Africa}\\
\mbox{$^2$ National Institute of Theoretical Physics, South Africa}
\mbox{$^3$National Laser Centre, Council for Scientific and Industrial Research, Pretoria, South Africa}}

\begin{abstract}
Unsharp POVM measurements allow the estimation and tracking of quantum wavefunctions in real-time with minimal
disruption of the dynamics. Here we demonstrate that high fidelity
state monitoring, and hence quantum control,  is possible even in the presence of classical
dephasing and amplitude noise, by simulating such measurements on a two-level system undergoing Rabi oscillations. 
Finite estimation fidelity is
found to persist indefinitely long after the decoherence times set by the noise fields in the absence
of measurement.
\end{abstract}

\maketitle

Maintaining high-fidelity quantum control is a central requirement in a variety of technologies ranging from nuclear
magnetic resonance to quantum based precision measurement
\cite{Chou2010}. Quantum control is usually restricted to a finite time window as a result of the unavoidable influence
of decohering environments, and the control lifetime is often extended through the use of decoherence free subspaces
\cite{Palma0396}, or dynamical decoupling \cite{Cywinski0508}.

In this communication we discuss a scheme which relies on a sequence
of consecutive POVM (Positive Operator-Valued Measure) measurements
to maintain quantum control.  We demonstrate that the time-evolution of a driven, isolated two-level quantum system,
subject to
classical dephasing and amplitude noise, can be monitored long beyond its Rabi coherence time. In fact, the wavefunction
can in principle be tracked indefinitely with finite fidelity, unlike systems controlled by dynamical decoupling that
ultimately undergo complete loss of coherence.  The control scheme
relies on periodic application of special POVM measurements said to be ``unsharp''
\cite{BuschGrabowskiLahti95}.  Such measurements have previously been shown to
allow faithful monitoring of Rabi oscillations  if the
general form of a time-independent Hamiltonian is known
\cite{Audretsch0401,Audretsch0107} and no external noise is present.

 A scheme for updating a state estimate during continuous measurements \cite{Belavkin89} -  the
continuum limit of the technique employed here -  has been presented in \cite{DiosiKonrad06}. In that scheme the
state evolution of the system, its estimated state, and the measurement readout is described by three
coupled stochastic differential equations, which indicate that the
state estimate converges to the real state for a broad
class of systems, cf.~\cite{KonradRothe10}.  Continuous measurements
have also been shown to drive statistical mixtures of
spatial wavepackets into pure states, which can be entirely determined by the measurement
record alone \cite{Doherty.et.al99}.

Specific experimental implementations of unsharp measurements have been
suggested in the context of Bose-Einstein Condensates
\cite{CorneyMilburn98}, cavity QED \cite{Audretsch0202} and coupled quantum dots
\cite{Korotkov03,OxtobyGambetta08}.  
Several realizations of the related topic of ``weak-value'' measurement have been demonstrated through measurements of
photon momentum \cite{RitchieStory1991, DixonStarling2009,HostenKwiat2008,KocsisBraverman2011}. In addition, experiments
of ``continuous weak measurement'' were implemented using a cold cesium vapor \cite{Silberfarb0705}.  
These realizations all employed ensemble measurements, while we here show monitoring of
a single, isolated quantum system by repeated measurement, as the system evolves.

We consider a two-level system undergoing Rabi-oscillations.  In a
frame rotating at the two-level transition frequency it evolves under the Hamiltonian
\begin{equation}
H_R = \hbar\frac{\Omega_R}{2}\hat\sigma_x,\label{Hrabi}
\end{equation}
where $\hat\sigma_x$ is the Pauli matrix which generates rotations about the $x$-axis and $\Omega_R$ the Rabi frequency,
which is assumed to be known. 
At the same time we assume the system is under the influence of random classical noise fields $\beta(t)$ and
$\alpha(t)$,
causing dephasing and amplitude fluctuations, respectively, through a noise Hamiltonian
\begin{equation}
H_N = \hbar\beta(t)\hat\sigma_z + \hbar\alpha(t)\hat\sigma_x.\label{Hnoise}
\end{equation}
Each noise field is characterized by a power spectrum which is related to its autocorrelation function
\begin{equation}
C^{(2)}(\tau) =  \frac{1}{T}\int_0^T \xi(t)\xi(t+\tau) dt \label{autoc}
\end{equation}
through
\begin{equation}
P_\xi(\omega)=\int C^{(2)}(\tau)e^{i\omega\tau}d\tau \label{pspec}
\end{equation}
where $\xi(t) = \alpha(t), \beta(t)$.

The estimation strategy rests on carrying out POVM measurements periodically \cite{NielsenChuang}, and updating the
state estimate based
on the measurement outcomes. Quite generally, a POVM measurement with outcome $n$, which was carried out on a system in
the state $|\psi\rangle$, will result in a state after the measurement given by
\begin{equation}
|\psi_n\rangle = \frac{\hat M_n|\psi\rangle}{\sqrt{p(n|\psi)}}.
\label{Kraus}
\end{equation}
Here $\hat M_n$ is the so-called ``Kraus operator'' corresponding to the measurement outcome $n$, and 
\begin{equation}
p(n|\psi) = \langle\psi|\hat M^\dagger_n\hat M_n|\psi\rangle\label{probn}
\end{equation}
is the probability to detect outcome $n$, conditioned on the system being in state $|\psi\rangle$.

In an estimation experiment a sequence of periodic measurements, with period $\tau$, are applied to the system as it
evolves in time \cite{Audretsch0401}. Despite the dynamics, the state change due to the
measurement can still be described by Eq.~(\ref{Kraus}) 
if each measurement is executed much faster than all other dynamical
timescales (impulsive measurement approximation).  In between measurements the time evolution is described
by the operator 
\begin{equation}
\hat U_j = \mathcal{T}\left[\exp{(-\frac{i}{\hbar}\int_{t_j}^{t_j+\tau} (H_R+H_N(t))dt)}\right],\label{timeevolv}
\end{equation}
where $\mathcal{T}$ is the time-ordering operator.  At $t=N\tau$, after $N$ measurements, the system is, up to the
appropriate
normalization constant, in the state
\begin{equation}
|\psi(N\tau)\rangle =\hat M_{n_N}\hat U_N\hat M_{n_{N-1}}\hat U_{N-1}...\hat M_{n_1}\hat
U_1|\psi\rangle.\label{finalstate}
\end{equation}
To estimate the state of the system the same sequence of operators corresponding to the measured outcomes in
Eq.~(\ref{finalstate}) are applied to an initial guess
$|\psi_e\rangle$ (cp.~\cite{DiosiKonrad06}), but \\
(1) the initial estimate of the state $|\psi_e\rangle$ can be taken as an arbitrary state vector on the Bloch sphere and \\
(2) in between measurements the state estimate is assumed to evolve only through the Hamiltonian Eq.~(\ref{Hrabi}),
since the experimenter does not know what the instantaneous values of
the noise fields are.  
As we'll see in what
follows it is still possible to estimate the state of the system without detailed knowledge of the noise fields. 
Our approach differs from \cite{Audretsch0107} where the Rabi
frequency was assumed to be unknown and one of the aims
was to determine
its value through a Bayesian estimator in the absence of noise. 

We now define two projectors
$\hat P_+ = \frac{1}{2}(\mathbb{1}+\hat{\mathbf r}\cdot\hat\sigma)$ and
$ \hat P_- =\frac{1}{2}(\mathbb{1}-\hat{\mathbf r}\cdot\hat\sigma)$,
where $\mathbb{1}$ is the identity operator, $\hat{\mathbf r} = (\delta,\zeta,\chi)$ a unit vector on the Bloch
sphere, and $\hat\sigma = (\hat\sigma_x,\hat\sigma_y,\hat\sigma_z)$. In terms of these it is possible to construct POVM
measurement operators
$\hat M_0 = \sqrt{p_0}\;\hat P_+ + \sqrt{1-p_0}\;\hat P_-$, and
$\hat M_1 = \sqrt{1-p_0}\;\hat P_+ + \sqrt{p_0}\;\hat P_-$,
related via
$M_0^\dagger M_0 +  M_1^\dagger M_1 = \mathbb{1}$,
and $0\leq p_0\leq0.5$.  The strength of a single measurement is quantified by $\Delta p= (1-p_0)-p_0= 1-2p_0$
\cite{AudretschDiosiKonrad02}. However, the strength of a sequence of
measurements depends also on the period $\tau$ between
two consecutive measurements. For fixed $\Delta p$ a shorter (longer)
period $\tau$ means a stronger (weaker) influence of the sequential
measurement. The strength of the state disturbance
due to this sequential measurement is best quantified by the
rate $\gamma_m=1/\tau_m$ with $\tau_m= 2\tau/(\Delta p)^2$  \cite{AudretschDiosiKonrad02}.
The strength $\gamma_m$ is the expected rate at which an arbitrary initial state is reduced to an
eigenstate of the measured observable, in the
absence of dynamics other than measurement \cite{Audretsch0401}.
\begin{figure}
\includegraphics[scale=0.45]{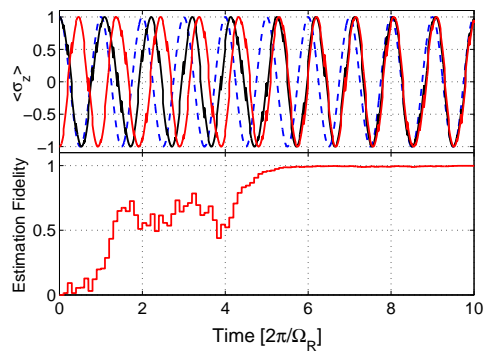}
\caption{Wavefunction estimation in the absence of noise for a specific realization of $\alpha(t)$ and $\beta(t)$
(single run). (a) Expectation value $\langle\hat\sigma_z\rangle$ for
blackline - true
expectation value, red line - estimated expectation value, dashed blue line - expectation value in
the absence of measurements, (b) Estimation fidelity.}\label{nonoisefig}
\end{figure}

To set the stage we illustrate the method in the absence of noise, i.e. $\beta(t)=\alpha(t)=0$.  We simulate an
experiment in which we choose $\Delta p = 0.2$ and $\hat{\mathbf{r}}=(0,0,1)$ which corresponds
to an unsharp measurement of $\hat\sigma_z$ (cp.\ \cite{Audretsch0107}). We
carry out a measurement every $\tau=T_R/10$, where $T_R=2\pi/\Omega_R$ is the Rabi-period.  
The resulting measurement strength is thus smaller than the
Rabi-frequency ($\gamma_m=\Omega_R/10\pi$), which is required in order not to
disturb the oscillations too strongly. For a measurement strength  $\gamma_m$
 much greater than $\Omega_R$ the state would be projected onto an eigenstate of the observable $\hat\sigma_z$ before a
single
oscillation can take place, and the dynamics would freeze (similarly to the Quantum Zeno Effect
\cite{MisraSudarshan77}). 
 The result of each
measurement is chosen at random, commensurately with the probabilities prescribed by Eq.~(\ref{probn}). The
initial state estimate is chosen orthogonal to the initial state vector, a limiting case for which the estimation
procedure might be expected to have some difficulty.  

In Fig.~\ref{nonoisefig}(a) we plot the expectation
value of $\langle\hat\sigma_z\rangle$  for the
true state (black line) and the state estimate (red
line) for one single run of the measurement experiment (i.e.~one specific realization of $\alpha(t)$ and $\beta(t)$). 
The state (black line) undergoes Rabi oscillations, but with measurement induced random phase shifts
as compared to the undisturbed oscillations  (dashed blue
line). The oscillations including the influence of
the measurement are monitored accurately by the estimate (red line) after about 6 Rabi
periods. After this time not only the expectation value of the measured
observable $\hat\sigma_z$  with respect to the true and estimated state,
but also the states themselves coincide as is
indicated by the plot of the estimation fidelity
$F(t)=|\langle\psi_{est}|\psi\rangle|^2$ in
Fig.~\ref{nonoisefig}(b). Asymptotically the fidelity tends to unity,
indicating the perfect state monitoring of a single system, in real
time, in the absence of noise.
\begin{figure}
\includegraphics[scale=0.45]{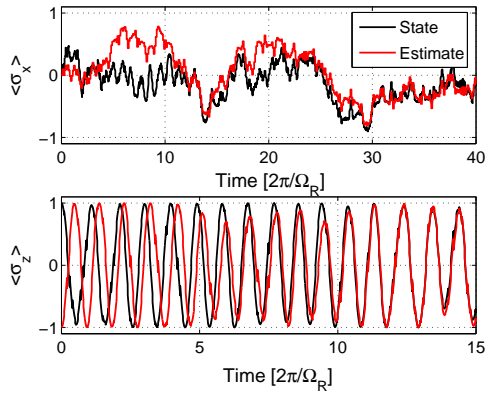}
\caption{State estimation in the presence of dephasing and amplitude noise.  (a) Expectation value of $\hat\sigma_x$ and
(b) expectation value of $\hat\sigma_z$ for a specific realization of $\alpha(t)$ and $\beta(t)$. Here we used
$\Delta\beta=0.05, \Delta\alpha=0.005$, $\Delta p = 0.2$, and
$\hat{\mathbf{r}}=(0.43,0,0.9)$. Note (a) and (b) have different time axes.}\label{withnoiseosc}
\end{figure}

Now consider a more realistic situation in which the two-level system is not isolated, but subject to random classical
noise as described by the Hamiltonian Eq.~(\ref{Hnoise}).  As an example we assume the noise fields both
have a power spectrum
$P_\xi(\omega)=A_\xi/\omega$, since this ``one-over-f" noise is ubiquitous in many systems.   For concreteness
we
choose
a lower cutoff of $\omega = 0.01$ and a high-frequency cutoff of $\omega=10$, where the frequency is specified in units
of $\Omega_R$.  In accordance with Eqs.~(\ref{autoc}) and (\ref{pspec}) we generate a specific noise trajectory
by summing over different spectral components, weighing each with the square root of the noise power:
$\xi(t) = \sum\limits_i \sqrt{P_\xi(\omega_i)} \cos{(\omega_i t + \phi_i)}.\label{pwsim}$
Each spectral component contains a random phase factor $\phi_i$, allowed to vary between $[0,2\pi]$, and assumed to be
delta correlated.
\begin{figure}
\includegraphics[scale=0.45]{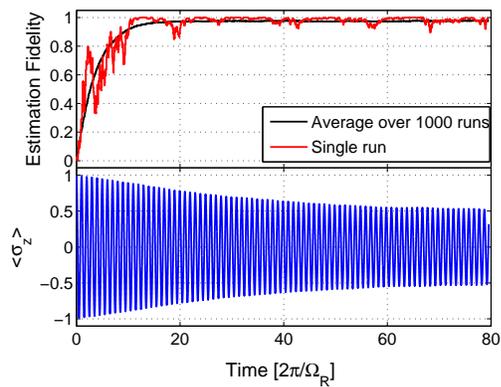}
\caption{State estimation in the presence of classical noise. (a) Red line - Estimation fidelity
for a single run (corresponding to Fig.~\ref{withnoiseosc}), Black line - expected
fidelity obtained by averaging 1000 runs.  (b) Blue line  - Rabi oscillations showing monotonic
loss of coherence in the absence of unsharp measurements.}\label{Rabivfidel}
\end{figure}
\begin{figure*}
\includegraphics[scale=0.5]{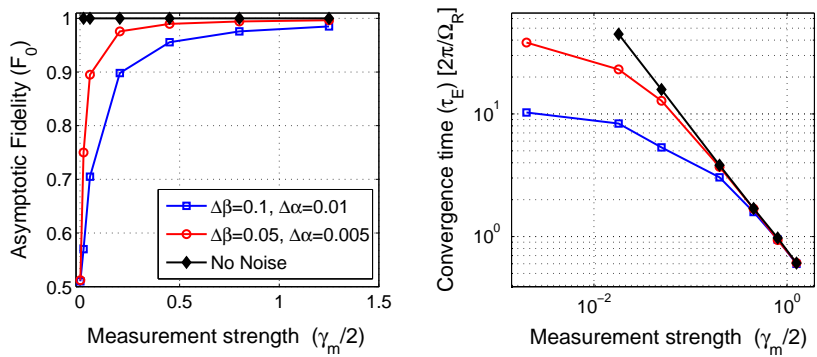}
\caption{Effect of measurement strength on state estimation in environments with different noise strengths: diamonds -
No noise, circles - $\Delta\beta=0.05, \Delta\alpha=0.005$, squares $\Delta\beta=0.1, \Delta\alpha=0.01$. (a)
Asymptotic fidelity, $F_0$, as a function of measurement strength.  (b)
Convergence time, $\tau_E$, as a function of measurement strength.  For the three weakest
$\gamma_m$'s on the curves with noise, we plot in (a) the asymptotic fidelity and in (b) that time which it
takes to reach a fraction $1-e^{-1}$ of that fidelity, even though the corresponding fidelity curves aren't
strictly exponential.}\label{strengthfig}
\end{figure*} 
Each noise trajectory, $\alpha(t)$ and $\beta(t)$, is normalized so that their root-mean-square deviations are
respectively one hundredth and one tenth of the drive field amplitude, $\Delta\alpha=0.005$ and
$\Delta\beta=0.05$. We use the same measurement strength as
  before, but an observable $\hat{\mathbf{r}}\cdot\hat{\sigma}$ with a finite x-
  component: $\hat{\mathbf{r}}=(0.43,0,0.9)$, since the noise is expected to tip the Bloch vector out of the
$yz$-plane.

Figure \ref{withnoiseosc} displays the evolution of the expectation values (a) $\langle\hat\sigma_x\rangle$ and
(b) $\langle\hat\sigma_z\rangle$ for the true state (black lines) and
the estimate (red lines), again for a single run of the experiment. The amplitude of the Rabi oscillations,
Fig.~\ref{withnoiseosc}(b), although
modulated by the noise, does not decrease permanently and the estimate succeeds
in tracking both components.
The red line in Fig.~\ref{Rabivfidel}(a) shows the estimation fidelity corresponding to the single run of the
experiment of Fig.~(\ref{withnoiseosc}).
Despite the noise the state estimate quickly approaches the real state, although the fidelity does not converge
completely to unity.  Instead, it exhibits random excursions away from unity, which at long times are centered
around an average, asymptotic value. To find this value we execute 1000 runs of the experiment with the same
initial conditions and average over the resulting fidelities, leading to the black curve in
Fig.~\ref{Rabivfidel}(a). We've found empirically that this average
fidelity $\bar{F}$ is  well described by 
\begin{equation}
 \bar{F}(t) = F_0(1-e^{-t/\tau_E}) \label{fidelf}.
\end{equation}
In Eq.~(\ref{fidelf}), $F_0$ is the asymptotic estimation fidelity and $\tau_E$
the estimation time. By fitting Eq.~(\ref{fidelf}) to the simulated result we extract an estimation time of
$\tau_E=3.7T_R$ and an asymptotic fidelity of $F_0=0.98$.
For comparison the blue curve in Fig.~\ref{Rabivfidel}(b) plots the result
of an average over 1000 simulated runs of the experiment in the \textit{absence of measurements}, but with the
same noise source as in (a). It shows the decay of Rabi oscillations to about half full amplitude over the same time span
due to the noise.  Labelling $\tau_{R}$ the characteristic decay time of Rabi oscillations, we remark that
Eq.~(\ref{fidelf}) holds accurately only when $\tau_m\ll\tau_{R}$.  For $\tau_m\gtrsim\tau_{R}$ the asymptotic approach
is no longer simply exponential.

The results of Figs.~\ref{Rabivfidel} (a) and (b) taken together imply that in any \textit{single run} of the
experiment the state can be estimated at all times after convergence with an average of 98\%
fidelity, by a pure state $|\psi_e(t)\rangle$.  On the other hand, in the absence
of measurements the state would lose coherence due to the noise, and evolve
into a statistical mixture as evidenced by the decay of
Rabi oscillations of the ensemble average shown in Fig.~\ref{Rabivfidel} (b).  This constitutes the main result of this
letter.
The fidelity can be operationally tested at the end of a run, by deducing from the state estimate the appropriate
unitary rotations needed to place the system in the state $|\!\uparrow\rangle$, say, where it
will then be detected with 98\% probability. As such, the experimenter has maintained quantum control by
monitoring of the state evolution, despite the noise.

Finally we study the effectiveness of the estimation process as a function of the measurement strength,
$\gamma_m$, when
noise is present.  The simulations are repeated for different values of $\gamma_m$,
and still choosing $\hat{\mathbf{r}}=(0.43,0,0.9)$ in each case. Figures \ref{strengthfig}(a)  and (b) respectively plot
the
estimation fidelities  and convergence times as a function of $\gamma_m$ for different noise strengths:
diamonds - no noise, circles - $\Delta\beta=0.05, \Delta\alpha=0.005$, squares - $\Delta\beta=0.1, \Delta\alpha=0.01$.
When noise is present, the asymptotic fidelity monotonically decreases as the measurement strength becomes weaker, and
approaches $F_0=0.5$ as $\gamma_m \rightarrow 0$. This is consistent with the average fidelity obtained when taking
random guesses for the state estimate. Simultaneously, the convergence time increases
as the measurement becomes weaker, but plateaus to a finite value as $\gamma_m\rightarrow 0$. By contrast,
in the absence of noise the fidelity always approaches $F_0=1$, but the convergence time increases indefinitely as the
measurement strength weakens, as can be seen in Fig.~\ref{strengthfig}(b).  

The trends observed 
in Fig.~\ref{strengthfig} emphasize that the appropriate timescales need to be obeyed for the measurement scheme to
work.  The sequential measurement must be weak enough not to freeze the dynamics, but strong enough to enable a high
fidelity estimate before the noise randomizes the system, i.e. $
T_R\ll\tau_m\ll\tau_R$. 

In conclusion, we remark that this study was carried out in a regime of comparatively strong noise, namely
$\Omega_R/\Delta\beta = 5\sim10$.  With stronger drive fields higher asymptotic fidelities can be expected for the same
measurement strengths considered here.  For example, we find that with $\Omega_R/\Delta\beta = 100$,
$\Omega_R/\Delta\alpha = 1000$, $\Delta p=0.1$,
$\hat{\mathbf{r}}=(0.43,0,0.9)$ ,that $F_0=0.999$ and $\tau=15.8 T_R$. It is encouraging that the 
estimation procedure described here predicts finite estimation fidelity despite the presence of random classical
noise. This opens the way for quantum control techniques that monitor wavefunction dynamics beyond the limitations set
by decoherence processes in the absence of unsharp measurements.


\end{document}